# On the existence of Cooper pairs in the Hubbard Model

Valentin Voroshilov

Physics Department, Boston University, 590 Commonwealth, Boston, MA 02215, valbu@bu.edu

By means of a new canonical transformation for the one-band Hubbard model at half filling we show the existence of Cooper pairs formed by strongly interacting quasiparticles.

71.10.Fd

74.20.Mn

One of the properties of HTSC[1] not yet fully understood is the existence of Cooper pairs at temperature higher than the critical temperature[2]. The Hubbard model attracts the attention of scientific community because of its possible connection to properties of HTSC[3]. A vast amount of literature is dedicated to the Hubbard model[4], in particular to studying its properties by means of a canonical transformation[5].

We start from a two-dimensional one-band Hubbard model[6] with Hamiltonian

$$H = -t \sum_{r,l,\sigma} (a^+_{r+l,\sigma} a_{r,\sigma} + a^+_{r,\sigma} a_{r+l,\sigma}) + U \sum_r a^+_{r,+} a_{r,+} a^+_{r,-} a_{r,-} . \qquad (1)$$

In Eq. (1) $t$ is a hopping integral for the nearest sites, $U$ describes on-site Coulomb repulsion, $r = (x, y)$ numerates sites of a square lattice, $l = (1,1)$ is a lattice vector; $a^+_{r,\sigma} (a_{r,\sigma})$ are creation (annihilation) Fermi-operators for an electron at the $r$-th site with spin projection $\sigma = \pm$ ½; and periodic boundary conditions are applied. For convenience, we set the lattice constant, the Boltzmann's constant, and the reduced Planck's constant equal to unity. We divide the lattice into two sublattices; let us call the sites "odd" when $x + y =$ odd number, and "even" otherwise.

We apply to the Hamiltonian (1) the canonical transformation (2) to new Fermi operators $b^+_{p,\sigma} (b_{p,\sigma})$

odd sites: $a_{r,+} = \dfrac{-1}{\sqrt{N}} \sum_p e^{ipr}(u_p b_{p,+} + v_p b^+_{-p,-})$; $\quad a_{r,-} = \dfrac{-1}{\sqrt{N}} \sum_p e^{-ipr}(u_p b^+_{p,-} - v_p b_{-p,+})$; (2a)

even sites: $a_{r,-} = \dfrac{1}{\sqrt{N}} \sum_p e^{ipr}(u_p b_{p,+} + v_p b^+_{-p,-})$; $\quad a_{r,+} = \dfrac{1}{\sqrt{N}} \sum_p e^{-ipr}(u_p b^+_{p,-} - v_p b_{-p,+})$; (2b)

$$b_{p,+} = \dfrac{-1}{\sqrt{N}} \sum_{r(odd)} e^{-ipr}(u_p a_{r,+} - v_p a_{r,-}) + \dfrac{1}{\sqrt{N}} \sum_{r(even)} e^{-ipr}(u_p a_{r,-} - v_p a_{r,+}); \quad (2c)$$

$$b_{p,-} = \dfrac{-1}{\sqrt{N}} \sum_{r(odd)} e^{-ipr}(v_p a^+_{r,+} + u_p a^+_{r,-}) + \dfrac{1}{\sqrt{N}} \sum_{r(even)} e^{-ipr}(v_p a^+_{r,-} + u_p a^+_{r,+}). \quad (2d)$$

In Eq. (2) $p = (p_x, p_y)$ and runs over all the values in the first Brillouin zone; $N$ is the number of sites; Hermitian conjugates are omitted.

Positive functions $u_p$ and $v_p$ are even in momentum space and satisfy the condition

$$u_p^2 + v_p^2 = 1. \quad (3)$$

When applying transformation (2) to the Hamiltonian (1), the result is

$$H = 4t \sum_p u_p v_p \omega_p - \dfrac{U}{N}\left(\sum_p u_p v_p\right)^2 + \dfrac{U}{N}\sum_{p_1 p_2} u_{p_1}^2 v_{p_2}^2 - 4t\sum_p u_p v_p \omega_p (n_{p,+} + n_{p,-}) +$$

$$+ \dfrac{U}{N}\sum_{p_1 p_2}\{(u_{p_1}u_{p_2} + v_{p_1}v_{p_2})^2 n_{p_1,+} - (v_{p_1}u_{p_2} - u_{p_1}v_{p_2})^2 n_{p_1,-} - (u_{p_1}u_{p_2} + v_{p_1}v_{p_2})^2 n_{p_1,+} n_{p_2,-} + \quad (4)$$

$$+ \tfrac{1}{2}(v_{p_1}u_{p_2} - u_{p_1}v_{p_2})^2 (n_{p_1,+}n_{p_2,+} + n_{p_1,-}n_{p_2,-})\} + V.$$

Term $V$ in Eq. (4) represents the part which is not diagonal with respect to $n_{p,\sigma}$, and $\omega_p = \cos p_x + \cos p_y$; $n_{p,\sigma} = b^+_{p,\sigma}b_{p,\sigma}$ in Eq. (4) and Eq. (5) but represents Fermi occupation numbers 0 or 1 in (6b).

Transformation (2) leads to the number of electrons

$$N_e = N + \sum_p (n_{p,+} - n_{p,-}). \quad (5)$$

To obtain self consistent equations for the energy spectrum and equilibrium distribution of quasiparticles we follow the well known variational procedure[7]; i.e., we neglect in Eq. (4) the non-diagonal term V and minimize the grand potential of the system, first with respect to $u_p$ and $v_p$ holding condition (3), and second with respect to $n_{p,\sigma}$ to define energy spectrum $\varepsilon_{p,\sigma}$.

For case $<n_{p,+}> = <n_{p,-}> = n_p$ (which is the only case we consider below) we automatically have a half-filling situation, $N_e = N$. Energy levels $E$, chemical potential $\mu$, energy spectrum $\varepsilon_p = \varepsilon_{p,\pm}$ and equations for functions $u_p$ and $v_p$ can be written as

$$p_{x,y} = \frac{2\pi m}{\sqrt{N}}; \quad m = 0, \pm 1, ..., \pm(\sqrt{N}-1)/2; \quad \mu = \frac{U}{2}; \quad \varepsilon_p = \frac{\Delta}{2(u_p^2 - v_p^2)}; \tag{6a}$$

$$E = C + \sum_p \varepsilon_p (n_{p,+} + n_{p,-}); \quad C = 4t\sum_p u_p v_p \omega_p - \frac{U}{N}\left(\sum_p u_p v_p\right)^2 + \frac{U}{N}\left(\sum_p u_p^2\right)\left(\sum_p v_p^2\right); \tag{6b}$$

$$(u_p^2 - v_p^2)\Omega_p = 2u_p v_p \Delta; \quad u_p^2 + v_p^2 = 1; \quad \Omega_p = -4t\omega_p + \Gamma; \tag{6c}$$

$$\Gamma = 2\frac{U}{N}\sum_p u_p v_p (1 - 2n_p); \quad \Delta = \frac{U}{N}\sum_p (u_p^2 - v_p^2)(1 - 2n_p). \tag{6d}$$

To satisfy equations and conditions (6a) – (6d) we have $\Gamma = 0$, $\Delta > 0$, and

$$u_p = \frac{1}{\sqrt{2}} * \begin{cases} \sqrt{1+X_p^2}; & \Omega_p > 0 \\ \sqrt{1-X_p^2}; & \Omega_p < 0 \end{cases}; \quad v_p = \frac{1}{\sqrt{2}} * \begin{cases} \sqrt{1-X_p^2}; & \Omega_p > 0 \\ \sqrt{1+X_p^2}; & \Omega_p < 0 \end{cases}; \quad X_p^2 = \frac{\Delta}{\sqrt{\Delta^2 + 16t^2\omega_p^2}}; \tag{7a}$$

$$\varepsilon_p = \frac{1}{2} * \begin{cases} \sqrt{\Delta^2 + 16t^2\omega_p^2}; & \Omega_p > 0 \\ -\sqrt{\Delta^2 + 16t^2\omega_p^2}; & \Omega_p < 0 \end{cases}; \quad \frac{U}{N}\sum_p \frac{\Omega_p}{|\Omega_p|} \frac{1-2n_p}{\sqrt{\Delta^2 + 16t^2\omega_p^2}} = 1. \tag{7b}$$

Apart from BCS theory[8] quasiparticles described by operators $b_{p,\sigma}^+ (b_{p,\sigma})$ exist at $T = 0$ ($n_p = 1$ for $\Omega_p < 0$ and 0 otherwise). The ground state energy includes the energy of quasiparticles for momenta with $\Omega_p < 0$;

$E_0 = C + 2\sum_{\Omega_p < 0} \varepsilon_p$; in a case of strong coupling $\Delta_0 \sim U >> 4t$ for the ground state energy we find $\frac{E_0}{Nt} \approx -\frac{U}{4t}$, which is in a qualitative agreement with other results[9]. The Fermi surface at a half filling is given by a square with the vertexes at $(0, \pm\pi)$ and $(\pm\pi, 0)$ ($\Omega_p < 0$ for the momenta inside of the square).

From Eq. (6) and (7) we find that parameter $\Gamma = 0$ at all temperatures, and for a strong coupling case with $\Delta \sim U >> 4t$ we obtain

$$\frac{\Delta}{U} \approx \tanh\frac{\Delta}{4T}. \tag{8}$$

Equation (8) has a solution only for $T < \frac{U}{4}$. When $T > \frac{U}{4}$, $\Delta = 0$ but quasiparticles still exist with $u_p = v_p = 1/\sqrt{2}$, and have a Fermi level at $\Omega_p = 0$.

The non-diagonal term V in Eq. (4) describes interactions between quasiparticles. Since quasiparticles have a Fermi level, if the interactions are effectively attractive, the ground state of the system has an instability relative to forming Cooper pairs between quasiparticles.

This last notion makes us reflect on similarities between quasiparticles in the Hubbard model and electrons in BCS model.

Let us assume that, like electrons in BCS theory, in the vicinity of the Fermi level quasiparticles experience effective attraction (this is the fundamental hypothesis for the following analysis). In this case we can write an effective Hubbard Hamiltonian essentially in the form of a BCS Hamiltonian with the energy spectrum $\varepsilon_p$ provided in Eq. (7b)

$$H_{eff} = E_0 + \sum_{\substack{\omega_p^2 < \lambda^2 \\ \sigma}} \varepsilon_p b_{p,\sigma}^+ b_{p,\sigma} - \frac{G}{N} \sum_{\substack{\omega_{p_1}^2 < \lambda^2 \\ \omega_{p_2}^2 < \lambda^2}} b_{p_1,+}^+ b_{-p_1,-}^+ b_{-p_2,-} b_{p_2,+} . \qquad (9)$$

with $0 < \lambda << \frac{\pi}{2}$, and $G > 0$.

The mathematics for Hamiltonian (9) is identical with the mathematics for BCS Hamiltonian, and we can write the new excitation spectrum $\Omega_p$ and the equation for gap $D$ as

$$\Omega_p = \sqrt{\Delta_*^2 + \varepsilon_p^2} = \sqrt{D^2 + 16t^2 \omega_p^2} ; \qquad m_p = \frac{1}{1 + \exp(\Omega_p / T)} ; \qquad \frac{G}{2N} \sum_{\omega_p^2 < \lambda^2} \frac{1 - 2m_p}{\sqrt{D^2 + 16t^2 \omega_p^2}} = 1 . \qquad (10)$$

In a case of a strong coupling between quasiparticles when $D \sim G\lambda/\pi >> 4t$, Eq. (10) leads to an equation

$$D \approx D_0 [1 - \frac{2}{1 + \exp(\frac{D}{T})}]; \qquad D_0 = \frac{G\lambda}{\pi} . \qquad (11)$$

For $T < \frac{U}{4}$, $\Delta \neq 0$, and solution (11) exists only if $D = \sqrt{\Delta_*^2 + \frac{1}{4}\Delta^2} > \Delta/2$. At $T \to 0$ condition $D > \Delta/2$ leads to condition $G > \frac{\pi U}{2\lambda}$, which means the effective interaction between quasiparticles has to be strong.

Equation (11) has a solution only for $T < \frac{\lambda G}{2\pi}$. When temperature is above $\frac{\lambda G}{2\pi}$ quasiparticles exist but do not form Cooper pairs. For large values of parameter $G$ and for temperatures $T < \frac{\lambda G}{2\pi}$ gap $D$ exists and strongly interacting quasiparticles (not weekly interacting electrons) form Cooper pairs. However, when temperature $T < \frac{U}{4}$ the second "gap" $\Delta \neq 0$ appears, which changes the relationship between electrons and quasiparticles ($u_p \neq v_p \neq 1/\sqrt{2}$). The physical nature of this evolution, its relevance to properties of HTSC (evolution of a "hidden Fermi liquid[10]"; transition to a superconductive state?), properties of the parameters of the model, and limits for the used approximations and hypothesis require furthermore investigation.

---

A simple and transparent model is presented with clear assumptions and apparent small parameters, which allows further generalization and verification. It is seen that ether the Hubbard model does not describe properties of HTCS (if similarly to the conventional superconductivity HTSC is a phenomenon of weekly interacting electrons), or superconductivity in HTSC should be described in terms of strongly interacting quasiparticles. Unfortunately, the term "quasiparticles" is rather ambiguous. When a theory starts from bare particles and than describes properties of a system in terms of dressed ones, those dressed particles/quasiparticles still have a close resemblance with the parent particles. Another situation is observed, for example, in BCS theory of superconductivity, where quasiparticles are actually "multi-particles", but still "constructed" from original (maybe even dressed) particles. The behavior of the system with Hamiltonian (1) appears to be better understood in terms of "multi-particles" constructed from "multi-particles" constructed from electrons.

[1] J. G. Bednorz and K. A. Muller, Z. Phys. B **64**, p. 189 (1986).

[2] Electron pairs precede high-temperature superconductivity; http://www.physorg.com/news145110552.html; Nov. 5 (2008)

[3] The theory of superconductivity in the high-Tc cuprates / P.W. Anderson. Princeton, N.J.: Princeton University Press, 1997, 446 p., pp. 20, 133.

[4] D. Baeriswyl, D. Campbell, J. M. Carmelo, F. Guinea, and E. Louis, The Hubbard model: Its Physics and Mathematical Physics (Springer, 1995); The Hubbard Model, edited by A. Montorsi, **Reprint** (World Scientific, 1992); Fabian H.L. Essler, Holger Frahm, and Frank Gohmann, The One-Dimensional Hubbard Model (Cambridge University Press, 2005).

[5] V. J. Emery, Phys. Rev. B **14**, p. 2992 (1976); Stellan Östlund, Phys. Rev. **69**, 1695-1698 (1992); I. E. Dzyaloshinskii, and A. I. Larkin, Sov. Phys. JETP **38**, 2002 (1974); Bang-Gui Liu, J. Phys.: Condens. Matter **6**, L415-L421 (1994); A. Dorneich, M. G. Zacher, C. Gröber, and R. Eder, arXive:cond-mat/9909351v1.; M. Cini, G. Stefanucci, and A. Balzarotti, Solid State Commun. **109**, 229-233 (1999); A. L. Alistratov, Yu. A. Dimashko, and V. S. Podol'skii, JETP Lett. **57**, 324-328 (1993); Florin D. Buzatu and Daniela Buzatu, Romanian Reports in Physics **59**, p. 351-356 (2007); F.D. Buzatu and G. Jackeli, Phys. Lett. A **46**, 163-171 (1998).

[6] J. Hubbard, Proc. R. Soc. A **276**, 238-257 (1963).

[7] N. N. Bogolubov, V. V. Tolmachev, and D. V. Shirkov, A New Method in the Theory of Superconductivity, Chapter 2, Appendix II (Consultants Bureau, New York, 1959); E. M. Lifshiz, and L. P. Pitaevskii, Statistical Physics, Part I I, Chapter 5 (Nauka, Moscow, 1978; Pergamon Press, 1980); Charles P. Poole Jr., Horacio A. Farach, and Richard J. Creswick, Superconductivity, p. 152-159 (Academic Press: New York, 1995); J. G. Valatin, Comments on the Theory of Superconductivity: in *The Theory of Superconductivity*: by N. N. Bogolubov, International Science Review Series, **Vol. 4**, pp. 118 – 132 (Taylor & Francis US, 1968).

[8] J. Bardeen, L. N. Cooper, and J. R. Schrieffer, Phys. Rev. 108, 1175 – 1204 (1957).

[9] Maciej M. Maska, Phys. Rev. B **57**, 8755 (1998).

[10] P. W. Anderson, arXiv:0709.0656v1 (cond-mat.str-ei)